\documentclass[conference]{IEEEtran}

\usepackage[super]{nth}
\usepackage{hyperref}
\usepackage{cite}
\usepackage[pdftex]{graphicx}
\usepackage{subcaption}
\usepackage{balance}

\newcommand{\FedComm}{\texttt{FedComm}}

\begin{document}

\title{\texttt{FedComm}: Understanding Communication Protocols for Edge-based Federated Learning}

\author{\IEEEauthorblockN{Gary Cleland\IEEEauthorrefmark{1}, Di Wu\IEEEauthorrefmark{2}, Rehmat Ullah\IEEEauthorrefmark{2}, and Blesson Varghese\IEEEauthorrefmark{1}\IEEEauthorrefmark{2}}
\IEEEauthorblockA{\IEEEauthorrefmark{1}\textit{School of Electronics, Electrical Engineering and Computer Science, 
Queen's University Belfast, UK}\\
}
\IEEEauthorblockA{\IEEEauthorrefmark{2}\textit{School of Computer Science, 
University of St Andrews, UK}\\
Corresponding author: bv6@st-andrews.ac.uk}
}
\maketitle
\thispagestyle{plain}
\pagestyle{plain}

\begin{abstract}
Federated learning (FL) trains machine learning (ML) models on devices using locally generated data and exchanges models without transferring raw data to a distant server. 
This exchange incurs a communication overhead and impacts the performance of FL training. There is limited understanding of how communication protocols specifically contribute to the performance of FL. Such an understanding is essential for selecting the right communication protocol when designing an FL system. This paper presents \FedComm\, a benchmarking methodology to quantify the impact of optimized application layer protocols, namely Message Queue Telemetry Transport (MQTT), Advanced Message Queuing Protocol (AMQP), and ZeroMQ Message Transport Protocol (ZMTP), and non-optimized application layer protocols, namely as TCP and UDP, on the performance of FL. \FedComm\ measures the overall performance of FL in terms of communication time and accuracy under varying computational and network stress and packet loss rates. Experiments on a lab-based testbed demonstrate that TCP outperforms UDP as a non-optimized application layer protocol with higher accuracy and shorter communication times for 4G and Wi-Fi networks. Optimized application layer protocols such as AMQP, MQTT, and ZMTP outperformed non-optimized application layer protocols in most network conditions, resulting in a 2.5x reduction in communication time compared to TCP while maintaining accuracy. The experimental results enable us to highlight a number of open research issues for further investigation. \FedComm\ is available for download from \url{https://github.com/qub-blesson/FedComm}.


\end{abstract}


\IEEEpeerreviewmaketitle

\section{Introduction}
\label{sec:introduction}
With billions of devices getting connected to the Internet, it is anticipated that 180 zettabytes of data will be generated by 2025~\cite{zwolenski2014digital}. The premise of edge computing is to (pre-)process data closer to the source (near to the devices) where it is generated. 

Machine learning (ML) techniques, such as federated learning (FL) have emerged to preserve the privacy of data when it is processed. Unlike traditional machine learning techniques, FL does not require raw data generated on the devices to be sent to a server for processing. Instead, FL involves a number of participating devices that train local neural network models that are sent to a server (for example, located on the edge) to aggregate the models and generate a global model.  

FL requires extensive communication between the devices and the server that significantly contributes to the overall time taken to train the model. The cost of communication is usually a bottleneck when the number of devices and the size of the model increases~\cite{Singh2019DetailedCO} and when there are network bandwidth constraints and high dimensional model updates~\cite{Asad2021EvaluatingTC}. Therefore, it is essential to optimise communication in FL.  

There is a significant body of research on strategies for optimising communication between devices and the server in FL~\cite{zhou2022you,mcmahan2017communication,chen2021communication}. 
Many of these strategies are highly intrusive and require substantial reengineering of the application since they are directly implemented at the algorithmic level of FL.

We argue that one key opportunity where there is a limited research focus for optimising communication in FL is the underlying communication protocols (both application and transport layers) that contribute to communication in FL. 
There are a variety of communication protocols, such as the Message Queue Telemetry Transport (MQTT)~\cite{hillar2017mqtt}, Advanced Message Queuing Protocol (AMQP)~\cite{AMQP}, ZeroMQ Message Transport Protocol (ZMTP)~\cite{ZMTP} and socket implementations of TCP and UDP that are available for implementing FL. 

However, there is a limited understanding of how these protocols impact communication in FL.
Such an understanding is essential to determine whether the communication protocols need to be carefully or can be arbitrarily chosen. If the protocols impact communication, then end-to-end system designers and developers will need to identify and select performance efficient combinations of application and transport layer protocols when developing novel edge computing systems and applications. 

This paper presents a methodology, namely \FedComm\ that benchmarks a variety of lightweight application layer and transport layer protocols within the context of FL. The classic FL implementation suitable for resource constrained devices and can operate in a device-edge environment presented in FedAdapt~\cite{wu2021fedadapt} is chosen for \FedComm.\ Metrics that capture performance aspects, including the accuracy of the model, and communication time in diverse network conditions, namely 3G, 4G and Wi-Fi based networks are considered. 

We anticipate that \FedComm\ will be useful to determine the answers of at least three fundamental questions: (i) How computational and network constraints, including packet loss affect the communication time and model accuracy of FL when using a given application layer protocol? (ii) Which application layer protocols are suitable for executing FL in resource constrained environments without requiring substantial or additional hardware and complex software environments? (iii) Given no constraints on the network and computation, which application layer protocol(s) deliver the best communication performance during FL training?

The results from a lab-based experimental testbed for \FedComm\ provides early insights into the effect of communication protocols on FL. 
The experimental studies highlight that when the network conditions are good (4G and Wi-Fi settings), TCP socket implementation achieves better communication time and accuracy than when using UDP. However, the TCP socket implementation poorly performs under poor network conditions in terms of communication time. In 4G and Wi-Fi networks, however, the effect on UDP accuracy is lower than that of TCP, but the communication time remains large. TCP accuracy is consistent even in poor network conditions, whereas UDP accuracy is significantly affected by poor network conditions such as 3G. TCP socket is considered the best non-optimized application layer protocol if the network conditions are not poor. Optimized application layer protocols, such as AMQP or MQTT, and ZMTP, produce the best results across all experiments, resulting in a communication time reduction of 2.5x compared to TCP socket while maintaining accuracy with a small difference of $<$0.5\%. \FedComm\ is available for public use\footnote{\url{https://github.com/qub-blesson/FedComm}}.

The contributions of this paper are: 

(i) The development of an FL benchmarking methodology, \FedComm\, that operates in a device-to-edge environment that quantifies the system performance when using different communication protocols. 

(ii) The evaluation of five application layer protocols that are relevant to FL in edge computing environments. 

(iii) The identification and collection of metrics that capture the performance of FL under varying network conditions using different application layer protocols. 

(iv) An experimental evaluation using \FedComm\ on a lab-based testbed. 

The rest of this paper is organised as follows. 
Section~\ref{sec:relatedwork} presents related research. 
Section~\ref{sec:benchmark} presents the \FedComm\ motivation, protocols used and the benchmarking methodology.
Section~\ref{sec:studies} presents the experimental studies from a lab-based testbed. 
Section~\ref{sec:discussion} concludes this paper by summarising the experimental studies and highlighting future work. 

\section{Related Work}
\label{sec:relatedwork}
This section will consider relevant research related to edge-based FL, communication overheads in FL and the communication protocols in FL.

\subsection{Edge-based FL}
Edge computing is a natural fit for applying FL paradigm as it facilitates collaborative training on end-user devices or edge nodes. There are two significant benefits in bringing together FL and edge computing: (i) user devices and edge nodes generate large volume of data, which need to be analyzed to extract insight from them for automated decision making by machine learning techniques~\cite{zhou2019edge}; (ii) FL provides a distributed machine learning solution to extract knowledge from data in a privacy-preserving manner~\cite{abreha2022federated}. 

However, running conventional FL in edge environments is challenging for two reasons: the relatively limited computational resources available on devices and a wide-range of heterogeneous devices~\cite{abreha2022federated,xia2021survey,khan2021federated}. In particular, end-user devices and edge nodes usually have limited computation resources compared to the cloud, resulting in impractical local training time in conventional FL~\cite{gao2020end,wang2020towards,das2019privacy}. In addition, end-user devices and edge nodes tend to be highly heterogeneous in terms of computation and communication capacity, which further degrades the performances of a edge-based FL system~\cite{imteaj2020federated,chai2019towards,li2020federated}. 

To address the above challenges, recent research has proposed reducing the amount of computational workload on end-user devices using techniques such as model pruning~\cite{jiang2022model}, computation offloading~\cite{thapa2022splitfed}, partial training~\cite{xu2021helios} and gradient sparsification~\cite{han2020adaptive}.
Personalized FL has also been proposed to surmount the challenges arising due to device heterogeneity in a cloud-edge based FL system~\cite{fallah2020personalized}. FedAdapt~\cite{wu2022fedadapt} adopts the reinforcement learning technique to adjust the amount of computation that is offloaded to the edge server, thereby adapting to varying resource availability and computational capacities of heterogeneous end-user devices. However, above researches rarely consider the effect of different communication protocols for a edge-based FL system since end-user devices and edge nodes always need customized communication protocols to meet their application requirements.

\subsection{Communication overhead in FL}
The communication overhead incurred in FL is well documented in the existing research literature~\cite{kairouz2021advances,konevcny2016federated,li2020federated}. Multiple communication-efficient approaches have been proposed to alleviate this problem. These approaches can be grouped into three categories - those that: (i) reduce the frequency of device updates, (ii) adopt compression schemes; (iii) optimise the FL architecture. 

There is research on selecting the participating end-user devices to reduce the overall communication costs between end-user devices and the server in each round of FL training~\cite{nishio2019client}. In addition, reducing the frequency of model updates for each device also reduces the total communication overhead~\cite{wang2019adaptive}. There is a large body of research that focuses on compression schemes. For example, FetchSGD compresses the gradient by employing a sketching technique~\cite{rothchild2020fetchsgd}. An online learning approach based on an adaptive degree of sparsity for gradient sparsification on non-i.i.d. local datasets is developed~\cite{han2020adaptive}. In addition, Fedpaq~\cite{reisizadeh2020fedpaq} introduces the quantization technique to compress the size of communicated updates of each end-user device before sending them to the server. 

In contrast to the above approaches, there is research that focuses on optimising the architecture of FL to reduce the overall communication overhead. RingFed~\cite{Yang2021RingFedRC} reduces communication costs by limiting data transfers between the server and devices. This is achieved by only one device receiving model updates from the server and then passing it on to other clients in a ring topology. HierFAVG~\cite{liu2020client} reduces the overall communication cost, training time and energy consumption of the end-user devices by introducing the intermediate edge servers in traditional cloud-based FL systems. 

Current communication reduction approaches in FL pay less attention to the basic communication protocols adopted between end-user devices and the server. Alternate approaches to the above as mentioned previously is to employ lightweight communication protocols, such as MQTT, AMQP, ZMPT and UDP, or develop new protocols that are suited for FL. However, the current literature has limited research on the impact of these protocols on FL.

\subsection{Communication Protocols in FL}
In an edge-based FL system, the global model is collaboratively trained across a large number of end-user devices with limited network bandwidth and unstable connections~\cite{wu2022fedadapt}. As a result, the communication cost is a key consideration in FL training. Current FL research and frameworks adopt the TCP protocol for the communication between end-user devices and the server~\cite{vineethfederated}. One major advantage of using TCP protocol is that it guarantees complete transmission of model updates from each end-user devices. However, when the network connections are relatively poor, the use of TCP leads to large re-transmissions, which negatively impacts the training time.

Developing the underlying communication protocols suited for FL will make data transmission more efficient. For instance, recent research explores the use of UDP protocols for communication and shows that UDP-based protocols will require less time than those that are TCP-based~\cite{vineethfederated}. In addition, a soft-DSGD~\cite{ye2022decentralized} method is proposed to mitigate the effect of missing model parameters (due to using UDP protocol). This is achieved by adjusting the weights of updated models during aggregation based on communication reliability. However, there is limited research on the underlying protocols and current research has neither systematically explored the TCP and UDP protocols nor have considered the protocols of the application layer. Therefore, the focus of this paper is a preliminary step in this direction to quantify the impact of communication protocols in FL.

\section{FedComm}
\label{sec:benchmark}
This section describes the rationale behind \FedComm\ benchmarking and its potential benefits and the communication protocols that are evaluated.

\subsection{Motivation}
To the best of our knowledge, benchmarking lightweight application layer protocols under FL has not been carried done to this extent before. The main goal of this study is to highlight the outcomes of evaluating lightweight application layer protocols in comparison to standard socket implementations of TCP. The emphasis is on identifying protocols that reduce the communication time and highlighting their pros and cons under different network conditions.



\texttt{FedComm} offers a series of tests that determines which communication protocols are beneficial in real-world FL scenarios. It should be noted that the aim of this paper is not to accelerate FL by implementing new mechanisms or protocols; rather, it considers existing application and transport layer protocols that developers can choose to easily incorporate into their own FL applications.

We anticipate a better understanding of application layer protocols in the context of FL as a result of this study. \texttt{FedComm} also benefits the FL community by utilizing lightweight application layer protocols to improve communication time without the use of additional hardware, specialized equipment, or large software environments.

\subsection{Protocols}
\begin{figure}[t]
	\centering
	\includegraphics[width=0.46\textwidth]{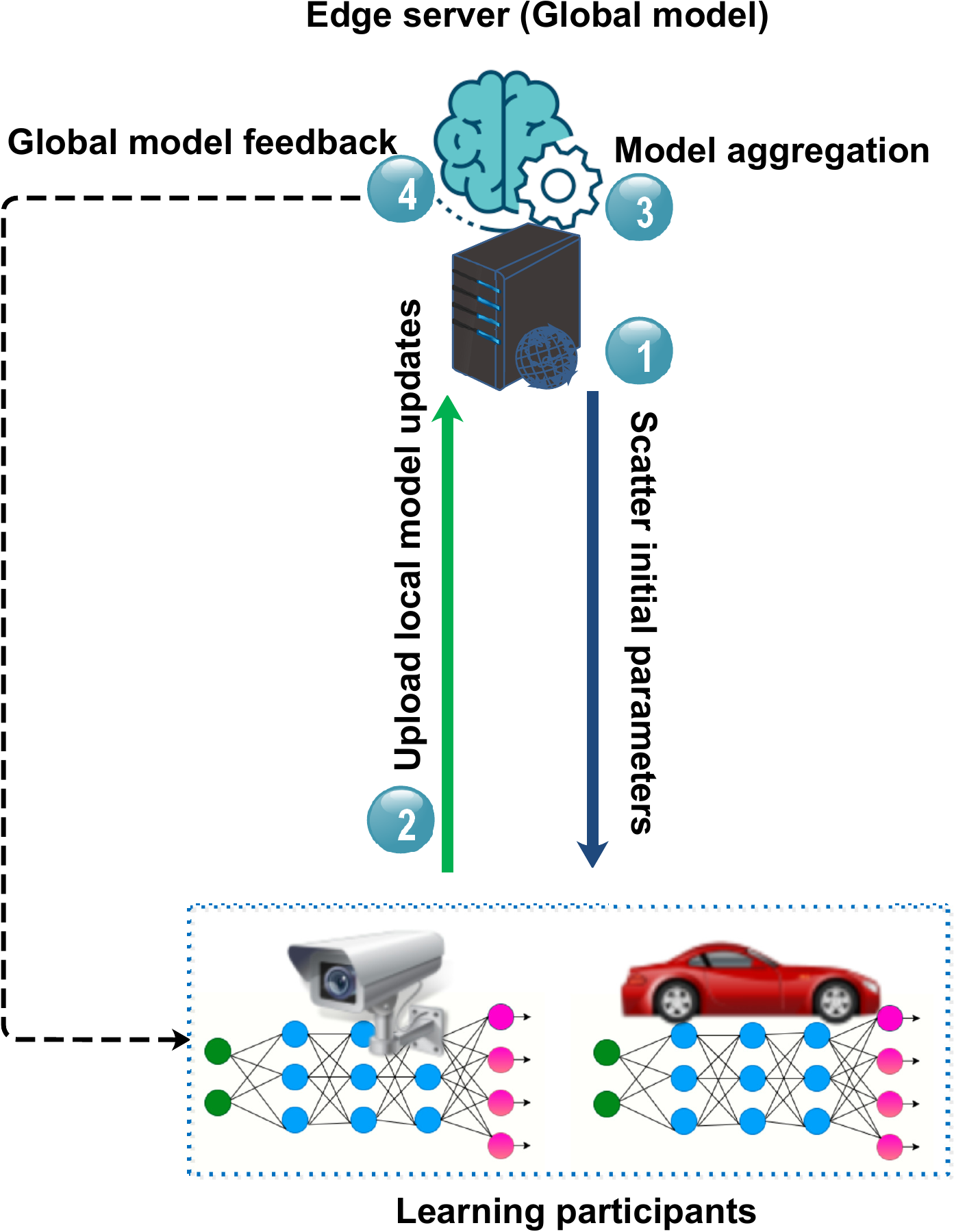}
	\caption{Overview of federated learning.}
	\label{fig:FL}
\end{figure}

Google \cite{konevcny2016federated} coined the term FL in 2016, which is a privacy-preserving ML technique in which an ML model is collaboratively learned across several distributed devices (e.g., mobile phones), while all training data is kept on local devices. An FL system is shown in Figure~\ref{fig:FL}. In the first step, the edge server initiates the global model and distributes the model parameters to all connected edge devices. In the second step, each local edge device trains a local version of the ML model on its local data. Instead of sending raw data to the server (edge), only the local model parameter updates are sent up to the server. In the third step, the server computes a weighted average of the parameter updates on the server using the Federated Averaging (FedAvg)~\cite{FedAvg} algorithm to obtain the new set of parameters for the global model. In the fourth step, a new global model is then sent back down to each device for the next round of training by the edge server. The entire  process is repeated until the model converges.

FL employs transport layer and application layer protocols to send and receive deep neural network (DNN) model parameters between the devices and the server. 
The developer can choose from protocols at two main layers of the OSI model: Layer 4 (transport) and Layer 7 (application). The developer must select a transport layer protocol that will serve as the foundation of communication. The developer then has the option of using a standard socket implementation of TCP/UDP or an alternative application layer protocol, such as MQTT, ZMTP, and AMQP.

At the application layer, {\texttt{FedComm}} currently evaluates socket TCP/UDP, MQTT, AMQP, and ZMTP. At the transport layer, all of these protocols use either TCP or UDP. The subsequent section discusses the transport layer and application layer protocols for the FL system.

\subsubsection{Transport Layer Protocols}

TCP as a transport layer protocol, is widely used in FL. TCP provides reliability through packet re-transmission, ensuring that model parameters are correctly sent and received by the device or server. TCP may be slower than other transport layer protocols, such as UDP, due to packet re-transmission. However, in terms of reliability, TCP remains the best transport layer protocol and is extensively used.

UDP on the other hand, is used in scenarios where packet loss is not a major concern and the overall performance measured by time is the most important factor. Packet loss can significantly affect model accuracy in the context of FL. Moreover, FL typically employs low-power devices, which are vulnerable to large packet loss rates under poor network conditions, limiting the learning rate of DNN models in comparison to TCP.

In {\texttt{FedComm}}, we evaluate both the above transport layer protocols using a standard socket implementation.

\subsubsection{Application Layer Protocols}
Application layer protocols are implemented at the seventh layer of the OSI model and are used for application communication. They provide rules for how data is exchanged. Source and destination applications must use the same protocol for accurate communication and to avoid transferring incompatible or inaccurate data that causes applications to fail. Application layer protocols have their own rules and can be used for purposes other than transferring data, such as transferring data in poor network conditions or providing lightweight communication for low power devices seen at the edge (or extreme edge), such as single board computers. 

FL algorithms with the proliferation of the Internet-of-Things are required to run on low-power devices that have relatively less processing power and will operate in `in-the-wild' scenarios that may have poor network conditions when compared to cloud data centre type environments. Application layer protocols designed for low-power IoT devices in poor network conditions, such as MQTT, AMQP, and ZMTP are useful in such cases for communication in FL.

\textit{MQTT} is a lightweight application layer protocol that implements a publish-subscribe architecture over TCP and is intended for use in low-power devices and microcontrollers in unreliable networks with limited bandwidth. There is no direct connection between a sender (publisher) and receiver (subscriber). An MQTT broker is used as a central system for communication between them and is responsible for filtering all incoming messages from publisher and distributing them to subscribers.
An MQTT client sends a single packet to the MQTT broker, which publishes it to all subscribers on devices and server backend systems. MQTT keeps a persistent connection to the broker.

\textit{AMQP} is an application layer protocol that runs over TCP at the transport layer and is designed for communication with middleware brokers such as RabbitMQ~\cite{10.5555/3126138}. AMQP is more advanced than MQTT, with built-in security and support for multiple architectures such as client/broker and client/server, and multiple exchange types such as direct, fan-out, topic, and headers. The AMQP fan-out exchange was used in \texttt{FedComm} because it routes messages to all queues that are bound to it, which is similar to what MQTT provides. Although AMQP has a larger header file than MQTT, it is faster and cost effective.


\textit{ZMTP} uses TCP at the transport layer and implements various architectures such as publish-subscribe, request-reply, and push-pull. ZMTP employs sockets rather than a broker server, potentially eliminating the broker overhead that may be present in other optimised application layer protocols.

\subsection{Benchmarking Methodology}

\begin{figure}[t]
	\centering
	\includegraphics[width=0.48\textwidth]{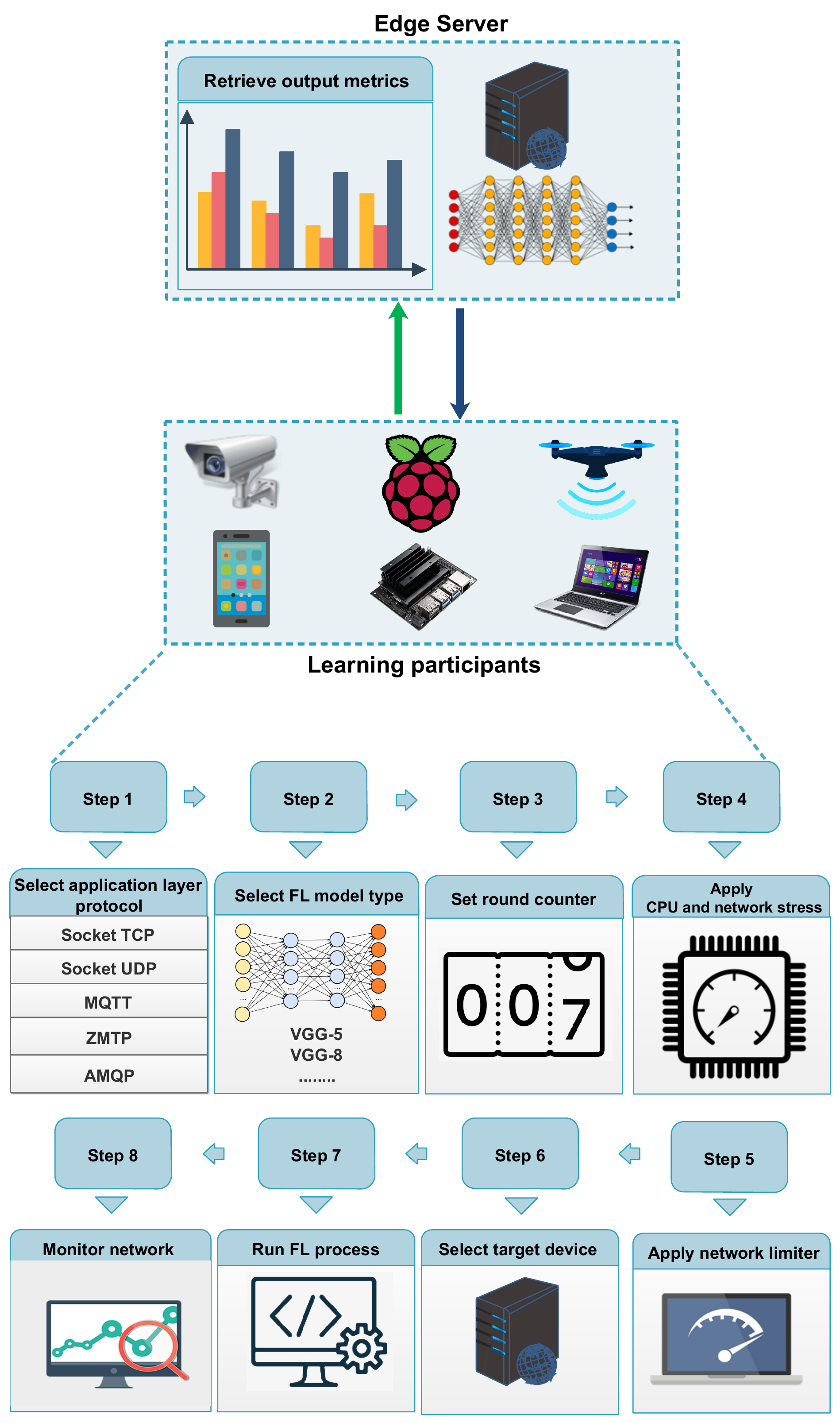}
	\caption{\FedComm\ benchmaking method.}
	\label{fig:fedbenchsteps}
\end{figure}

\texttt{FedComm} aims to automatically benchmark FL using a variety of application layer protocols, with a focus on communication time and accuracy. Figure~\ref{fig:fedbenchsteps} illustrates the \texttt{FedComm} method, which consists of the following steps:


\textit{\textbf{Step 1 - Select application layer protocol}}: 
In the first step, the user manually selects from the list of application\footnote{TCP socket and UDP socket are non-optimized application layer protocols, and MQTT, AMQP, and ZMTP are optimised application layer protocols.} layer protocols available in \texttt{FedComm}. All devices must communicate using the protocol selected at the application layer in order to transfer FL model parameters. This step is optional; if not chosen, then communication will take place via the TCP socket by default.

\textit{\textbf{Step 2 - Select the FL model type}}:
In the second step, the user chooses from a list of available FL models\footnote{Support for the VGG-5 and VGG-8 DNN models is currently offered. The implementation can be easily extended to support other models.}. The model will be transferred between devices and server using the application layer protocol chosen from Step 1. This step is optional; if no model is selected, then in \FedComm\ the VGG-8 model is used as the default.


\textit{\textbf{Step 3 - Set round counter}}:
In the third step, the user determines how many rounds to run FL on all devices. One round consists of all devices receiving a model with initial or new weights from the server. Devices then train the selected model locally and send new weights to the server, where the aggregation process combines the trained model weights. This is optional; if the counter is not specified, then \FedComm\ runs five FL rounds as default.

\textit{\textbf{Step 4 - Apply CPU and network stress}}:
CPU and network stress is applied in this step. CPU stress is applied by performing intensive matrix computations to rapidly increase CPU activity to reduce the amount of computational resources available for running FL to represent the environment of a device with relatively limited computational capabilities. Network stress is applied by sending an influx of TCP packets between the device and the server in order to generate network traffic for representing a networked device operating in an environment with varying network conditions. The user is provided with the option to either select CPU or network stress; this step is optional, and no stressor is selected by default.

\textit{\textbf{Step 5 - Apply network limiter}}:
Using an internal bandwidth limiter, this step emulates specific network conditions, similar to Step 4. A list of bandwidth limits are provided as options to the user to select from, including a 5Mbps limit for 3G bandwidth, a 20Mbps limit for 4G bandwidth, and a 60Mbps limit for Wi-Fi bandwidth. This step is optional, and by default, the entire available bandwidth is utilised and no network constraints are applied.

\textit{\textbf{Step 6 - Select target device}}:
In this step, the IP address of the server and of the participating devices are provided by the user as metadata in a configuration file. The FL process is initiated on the server. 

\textit{\textbf{Step 7 - Run FL process}}:
After completing all of the preceding steps, a classic FL task will be executed (as shown in Figure~\ref{fig:FL}. The server will initiate model parameters and distribute them to each participating device via the application layer protocol. The devices will train the model and send it back to the server for testing, where the server collects the necessary metrics and stores them as a file.

\textit{\textbf{Step 8 - Monitor network}}:
In this step, the network is monitored for the re-transmissions of TCP packets. 
Once the FL process is completed, the number of packets re-transmitted across all clients is recorded, and output metrics are saved to a pickle file on the edge server.



\section{Experimental Studies}
\label{sec:studies}
This section presents the evaluation results obtained using \FedComm\. The experimental setup and protocol setup are presented in Section~\ref{sec:setup}. Section~\ref{sec:metrics} discusses performance evaluation metrics, and Section~\ref{sec:results} presents the results.

\subsection{Setup}
\label{sec:setup}
\subsubsection*{Experimental Setup}
All experiments were carried out on a lab-based testbed  comprising five virtual machines running Ubuntu 20.04.3 LTS, one of which serves as the server and the other four as devices. The server has 16GB of DIMM DRAM EDO memory, an octa-core Intel Xeon E5 CPU E5-2695 v4 @ 2.10GHz, and 105GB of storage. The devices include 2GB of DRAM memory, a single core of the Intel Xeon E5 CPU E5-2695 v4 @ 2.10GHz, and 30GB of storage.

The DNN models used are VGG-5 and VGG-8, and the dataset is CIFAR-10~\cite{Krizhevsky2009LearningML}. The dataset contains 50K training samples and 10K testing samples. The FedAvg~\cite{FedAvg} aggregation method is used on the server.

\subsubsection*{Protocol Setup} 
The configuration related to enabling the protocols for \FedComm\ are as follows:

(i) TCP, UDP and ZMTP required no additional configuration; the built-in Python module socket was used for communication. 

(ii) Experiments using MQTT require the installation of Mosquitto~\cite{Light2017}, a MQTT broker, on the server machine where all clients and server establish a connection using the \texttt{paho-mqtt} Python module. To enable remote connections from other client machines, a new Mosquitto configuration file must be created with the listener set to port 1883 and anonymous connections enabled. The Mosquitto broker must be started before using MQTT within \FedComm.\

(iii) Similar to MQTT, AMQP experiments require the installation of the open source messaging broker RabbitMQ, and a connection must be established between all client machines and a single server in order to send and receive messages. The Python \texttt{pika} module is required for the client-server message exchange.

\subsection{Benchmark Metrics}
\label{sec:metrics}
The performance metrics for the evaluation of protocols used in \FedComm\ are as follows:

1) \textbf{Communication time} - This metric calculates the total communication time across all FL rounds. The communication time includes the time spent sending and receiving model parameters between the devices and server during FL training. We aim to understand the impact on communication time when computational stress and network stress are there in the environment and packet losses occur. 

a) \textit{Communication time under computational stress:} 
This metric measures the total communication time under computational stress. The \texttt{stress-ng}~\cite{stressng} module is used to spawn threads to increase the number of computations performed on the compute core of the device in order to simulate a working networked device. Stress-ng computes 830 ops/s using 99\% of the CPU power to simulate a low-power device incapable of performing large tasks.

b) \textit{Communication time under network stress:} 
This metric measures the total amount of communication time spent under network stress. \texttt{Netstress} emulates an active network, sending messages at around 250 Mbits/s.

c) \textit{Communication time under packet loss:} 
This metric measures the total amount of communication time spent when the packet loss rate varies. Packet loss is calculated as the percentage of packets received incorrectly or not at all at a destination. To simulate packet loss, we use the \texttt{tc} tool and simulate 0\%, 5\%, 10\%, and 20\% packet loss rates.

2) \textbf{Accuracy} - Accuracy is defined as the number of correct predictions divided by the total number of predictions. This metric measures how accurate the new aggregated model is against the test data in each FL round.

\subsection{Results}
\label{sec:results}
This section evaluates the non-optimized and optimized application layer protocols under varying network conditions and presents the above performance metrics.

1) \textit{Non-optimized Application Layer Protocols}

\begin{figure}[tp]
     \centering
     \begin{subfigure}[b]{0.24\textwidth}
         \centering
         \includegraphics[width=\textwidth]{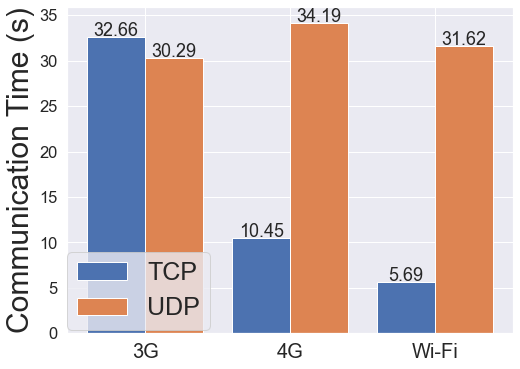}
         \caption{VGG-5}
    \label{fig:tcp_udp_comm_vgg5}
     \end{subfigure}
     \hfill
     \begin{subfigure}[b]{0.24\textwidth}
         \centering
         \includegraphics[width=\textwidth]{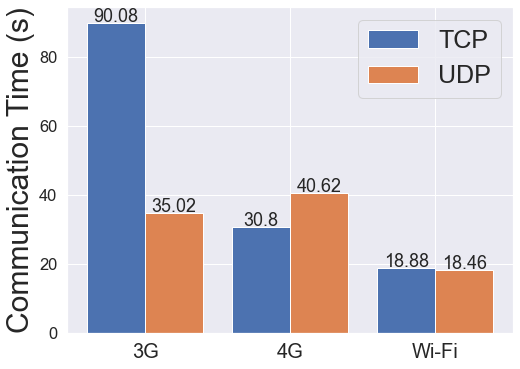}
         \caption{VGG-8}
    \label{fig:tcp_udp_comm_vgg8}
     \end{subfigure}
        \caption{Communication time when using TCP and UDP sockets after 5 training rounds for VGG-5 and VGG-8 models under different network settings when under no stress.}

\end{figure}

\textbf{Communication time without computational/network stress:} Figure~\ref{fig:tcp_udp_comm_vgg5} and Figure~\ref{fig:tcp_udp_comm_vgg8} shows the communication time by comparing TCP socket and UDP socket after 5 rounds of the VGG-5 and VGG-8 models for varying network settings. When the network bandwidth increases from that of 3G to Wi-Fi, there is a consistent decrease in communication time when using TCP for both VGG-5 and VGG-8. Moreover, when the model parameters are smaller, such as for VGG-5, TCP sockets significantly outperform UDP sockets in terms of time spent transferring the model to another device under 4G and Wi-Fi network conditions. However, when network conditions are relatively poorer, such as in 3G, UDP outperforms TCP and is suitable for larger models. TCP performs well when measuring communication time for relatively good network bandwidths and small models. When the network condition is poorer and large models are required, UDP performs better than TCP. This is because there are no re-transmissions due to packet loss in the network when using UDP. However, TCP by design re-transmits packets, increasing communication time if the network condition is poor.

\begin{figure}[tp]
     \centering
     \begin{subfigure}[b]{0.24\textwidth}
         \centering
         \includegraphics[width=\textwidth]{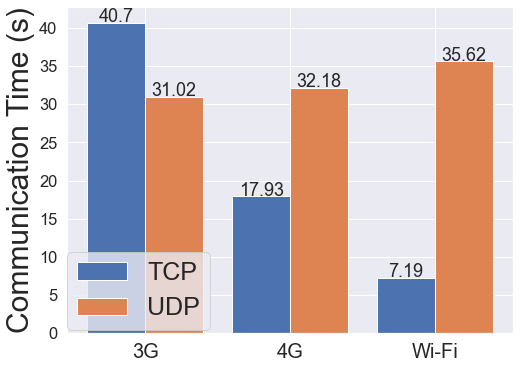}
         \caption{VGG-5}
    \label{fig:tcp_udp_cpu_vgg5}
     \end{subfigure}
     \hfill
     \begin{subfigure}[b]{0.24\textwidth}
         \centering
         \includegraphics[width=\textwidth]{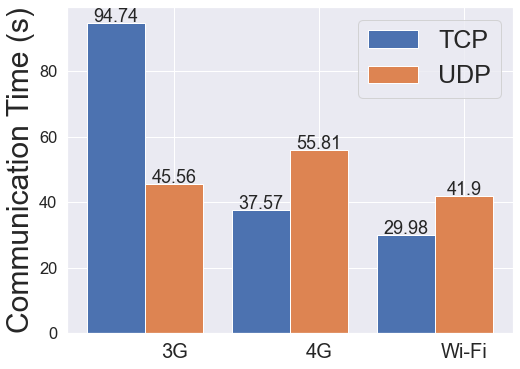}
         \caption{VGG-8}
    \label{fig:tcp_udp_cpu_vgg8}
     \end{subfigure}
\caption{Communication time when using TCP and UDP sockets after 5 training rounds for VGG-5 and VGG-8 models under different network settings while under CPU stress.}

\end{figure}

\textbf{Communication time under computational stress:} Figure~\ref{fig:tcp_udp_cpu_vgg5} shows that when 99\% CPU stress is applied, communication time using the TCP socket is affected more than UDP for the VGG-5 model. In comparison to Figure~\ref{fig:tcp_udp_comm_vgg5}, Figure~\ref{fig:tcp_udp_cpu_vgg5} shows that TCP increases by a total of 16.82 seconds across all network conditions, whereas UDP maintains consistency while only adding 2.72 seconds to total communication time. Figure~\ref{fig:tcp_udp_cpu_vgg8} shows that when CPU stress is applied to the larger VGG-8 model, the UDP communication time increases by 49.19 seconds, whereas the TCP socket communication time increases by 22.61 seconds when compared to Figure~\ref{fig:tcp_udp_comm_vgg8}. Overall, CPU stress has a significant impact on UDP for larger models. This is due to the implementation of UDP and the additional processing required to send and receive model parameters in FL. Computationally intensive DNN models will therefore have a substantial impact on UDP performance on low-powered devices with poor CPU performance.

\begin{figure}[tp]
     \centering
     \begin{subfigure}[b]{0.24\textwidth}
         \centering
         \includegraphics[width=\textwidth]{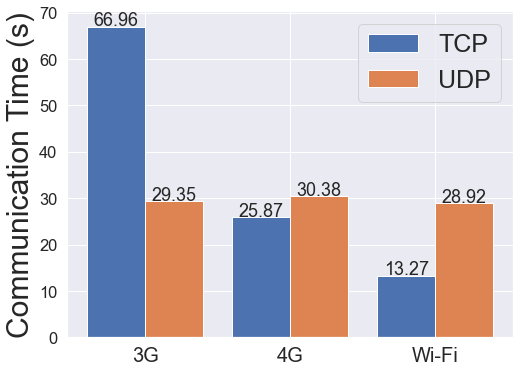}
    \caption{VGG-5}
    \label{fig:tcp_udp_net_vgg5}
        
     \end{subfigure}
     \hfill
     \begin{subfigure}[b]{0.24\textwidth}
         \centering
         \includegraphics[width=\textwidth]{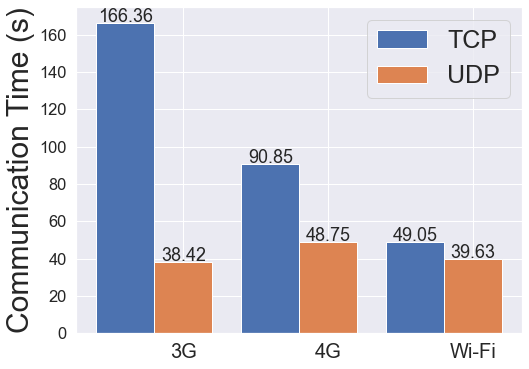}
    \caption{VGG-8}
    \label{fig:tcp_udp_net_vgg8}
     \end{subfigure}
\caption{Communication time when using TCP and UDP sockets after 5 training rounds for VGG-5 and VGG-8 models with different network settings under network stress.}
\end{figure}

\textbf{Communication time under network stress:}
Figure~\ref{fig:tcp_udp_net_vgg5} and Figure~\ref{fig:tcp_udp_net_vgg8} show the communication time when network stress is applied when using VGG-5 and VGG-8 models. 
It is noted that network stress influences UDP less adversely than CPU stress due to the additional computations performed during CPU stress. Moreover, the observed performance of UDP under network stress when using the VGG-5 model may also be since a small number of model parameters that have no effect on the overall communication time are used. As an application layer protocol, the TCP socket is significantly impacted by network stress, which increases communication time when compared to CPU stress. UDP outperforms TCP in all network stress tests conducted with the VGG-8 model. The results shown in Figure~\ref{fig:tcp_udp_net_vgg5} and Figure~\ref{fig:tcp_udp_net_vgg8} indicate that UDP performs well under 3G network conditions and can therefore be recommended if accuracy is not a concern.

\begin{figure}[tp]
     \centering
     \begin{subfigure}[b]{0.24\textwidth}
         \centering
         \includegraphics[width=\textwidth]{ 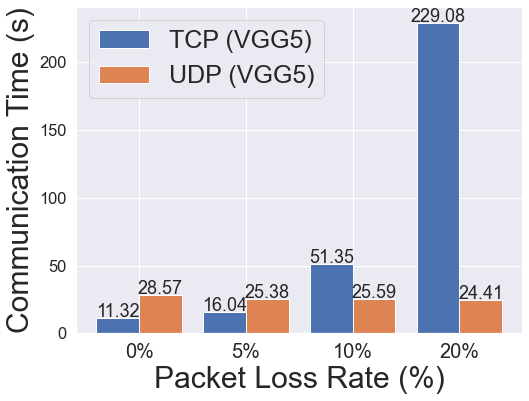}
\caption{VGG-5}
    \label{fig:tvu5}

     \end{subfigure}
     \hfill
     \begin{subfigure}[b]{0.24\textwidth}
         \centering
         \includegraphics[width=\textwidth]{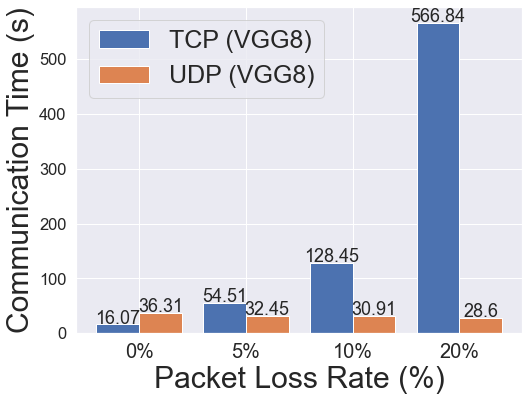}
         
    \caption{VGG-8}
    \label{fig:tvu8}
     \end{subfigure}
\caption{Communication time when using TCP and UDP sockets after 5 training rounds for VGG-5 and VGG-8 models under different packet loss rates.}
\end{figure}

\textbf{Communication time under packet loss rate:}
Figure~\ref{fig:tvu5} and Figure~\ref{fig:tvu8} shows the communication time under various packet loss rates of 0\%, 5\%, 10\%, and 20\% for both the VGG-5 and VGG-8 models. It is evident that UDP communication time is uniform across varying packet loss rates, whereas TCP communication time increases significantly when the packet loss ranges from 0\% to 20\%. Variable packet loss rates have a significant impact on TCP socket and the major reason is that TCP re-transmits all packets that were received incorrectly or were not received. A lower communication time is observed for UDP since it is a connectionless protocol that does not re-transmit packets.

\begin{figure}
	\centering
    \includegraphics[width=0.50\textwidth]{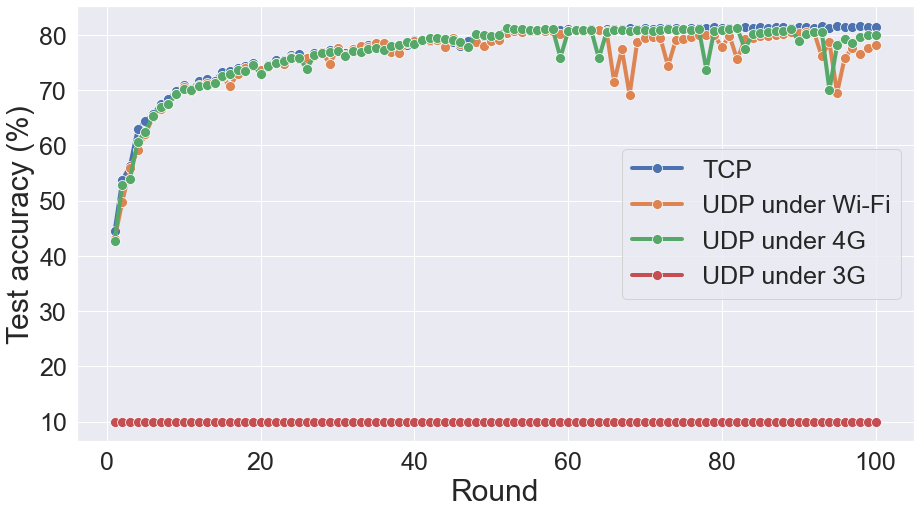}
	\caption{Accuracy when using TCP and UDP sockets after 100 training rounds of the VGG-5 model under different network conditions.}
\label{fig:tcp_udp_acc_vgg8}
\end{figure}

\textbf{Accuracy:}
Figure~\ref{fig:tcp_udp_acc_vgg8} highlights that TCP maintains accuracy in all network conditions because TCP is a reliable protocol that re-transmits all packets that were incorrectly received or were not successfully transmitted. TCP re-transmits packets until all packets are successfully received, and as a result, accuracy is preserved. On the other hand, the accuracy of UDP is inconsistent, particularly for 3G networks. When the network condition is poor, there is a greater possibility of packet loss, which impacts the overall accuracy. The packet loss rate for 4G and Wi-Fi network conditions is relatively low; there is a negligible decrease in accuracy when switching from 4G to Wi-Fi. It should be noted that there is no direct link between packet loss and the quality of network conditions, as other factors, such as network congestion, transmission power and fading can affect the rate at which packets are lost.

2) \textit{Non-optimized Application Layer Protocols}
When comparing the results, TCP socket is the preferred non-optimized application layer protocol in most cases, and all of the optimized application layer protocols tested in \FedComm\ use TCP at the transport layer. Therefore, this section compares the optimized application layer protocols when using the non-optimized socket implementation of TCP.

\begin{figure}[tp]
     \centering
     \begin{subfigure}[b]{0.48\textwidth}
         \centering
         \includegraphics[width=\textwidth]{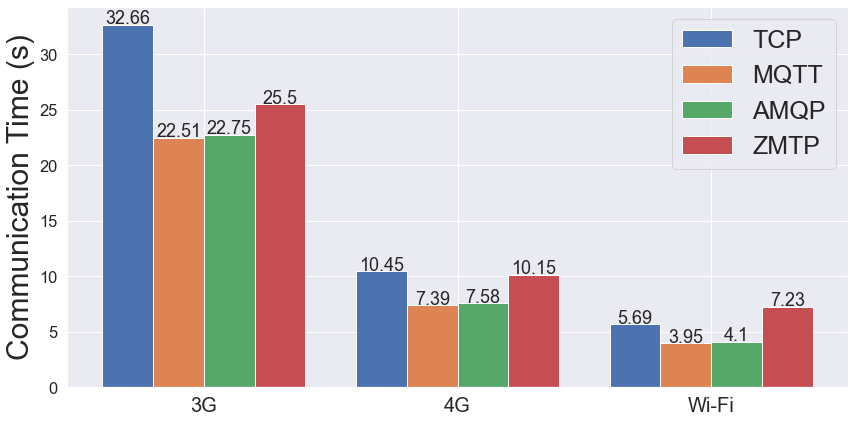}
        \caption{VGG-5}
    \label{fig:tcp_mqtt_amqp_comm_vgg5}
     \end{subfigure}
     \hfill
     \begin{subfigure}[b]{0.48\textwidth}
         \centering
         \includegraphics[width=\textwidth]{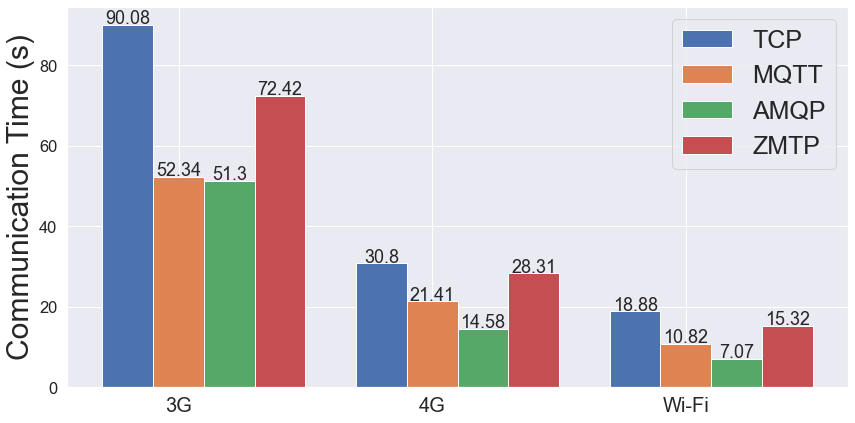}
        \caption{VGG-8}
    \label{fig:tcp_mqtt_amqp_comm_vgg8}
     \end{subfigure}
         \caption{Communication time when using TCP socket, MQTT, AMQP and ZMTP after 5 training rounds of the VGG-5 and VGG-8 model under different network settings.}
\end{figure}

\textbf{Communication time without computational/network stress:} Figure~\ref{fig:tcp_mqtt_amqp_comm_vgg5} and Figure~\ref{fig:tcp_mqtt_amqp_comm_vgg8} shows the communication time against various network conditions involving TCP and all application layer protocols under the VGG-5 and VGG-8 models. Under most network conditions, TCP's performance is poor resulting in the largest communication time. AMQP is 1.75x faster than TCP in 3G network conditions, 2.15x faster in 4G network conditions, and 2.5x faster in Wi-Fi. AMQP and MQTT perform almost the same in all network conditions with both models. This is potentially because AMQP and MQTT use brokers to distribute a single message to all devices, whereas TCP requires one message per device. ZMTP, despite being an optimized application layer protocol, performs similarly to TCP in most network conditions, as shown in Figure~\ref{fig:tcp_mqtt_amqp_comm_vgg5} and Figure~\ref{fig:tcp_mqtt_amqp_comm_vgg8}. MQTT performs better when small amounts of data are transferred (for the VGG-5 model), whereas AMQP performs better overall for the VGG-8 model when more data is transferred per round.

\begin{figure}[tp]
     \centering
     \begin{subfigure}[b]{0.48\textwidth}
         \centering
         \includegraphics[width=\textwidth]{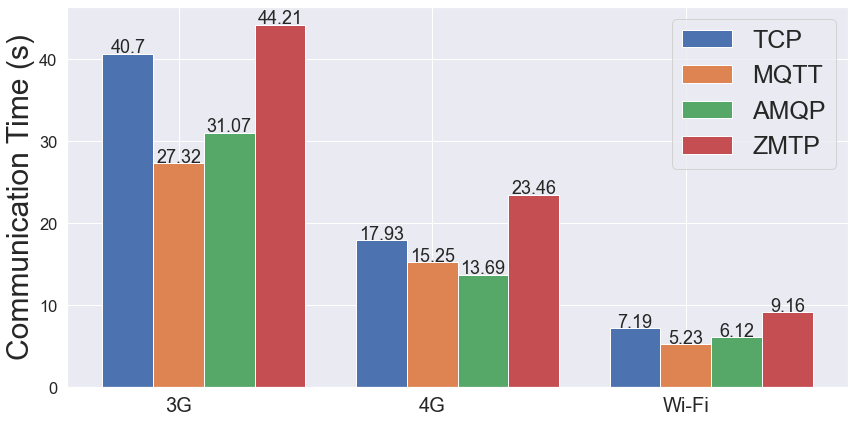}
        \caption{VGG-5}
    \label{fig:tcp_mqtt_amqp_cpu_vgg5}
     \end{subfigure}
     \hfill
     \begin{subfigure}[b]{0.48\textwidth}
         \centering
         \includegraphics[width=\textwidth]{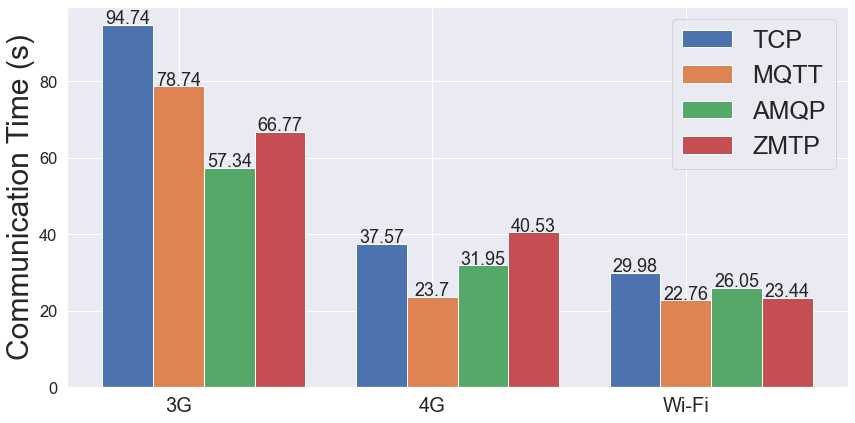}
        \caption{VGG-8}
    \label{fig:tcp_mqtt_amqp_cpu_vgg8}
     \end{subfigure}
         \caption{Communication time when using TCP socket, MQTT, AMQP and ZMTP after 5 training rounds of the VGG-5 and VGG-8 models for different network settings while under CPU stress.}
\end{figure}

\textbf{Communication time under computational stress:} CPU stress on all devices has different effects on each protocol because they are designed for use in different scenarios. With 99\% CPU stress, TCP has a 22.5 second increase in communication time, MQTT has a 40.6 second increase, and AMQP has a 42.3 second increase from Figure~\ref{fig:tcp_mqtt_amqp_comm_vgg8} to Figure \ref{fig:tcp_mqtt_amqp_cpu_vgg8}. The IoT-based protocols gain a larger increase in communication time, but they outperform socket TCP. The AMQP implementation relies on additional threads created by the developer to listen for incoming messages, which can be limiting for devices that can only run a single thread. This has a significant impact on computation, resulting in a large increase in communication time. MQTT is similar to AMQP, but handles the additional thread automatically. TCP does not require threads, which reduces the impact of CPU stress.

\begin{figure}[tp]
     \centering
     \begin{subfigure}[b]{0.48\textwidth}
         \centering
         \includegraphics[width=\textwidth]{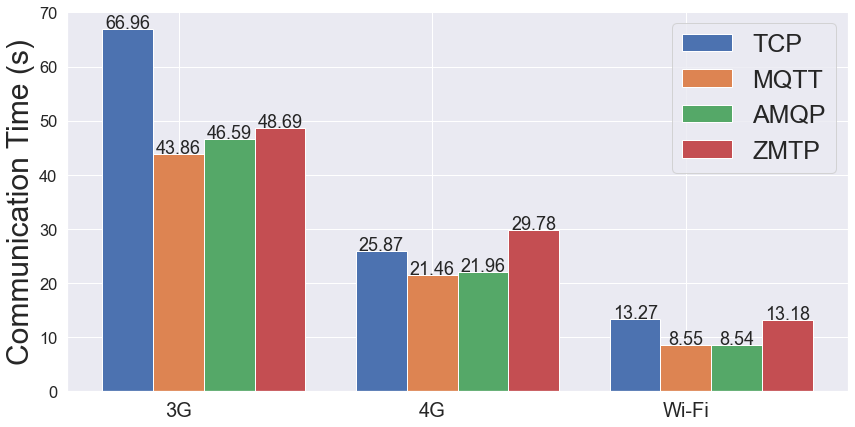}
        \caption{VGG-5}
    \label{fig:tcp_mqtt_amqp_net_vgg5}
     \end{subfigure}
     \hfill
     \begin{subfigure}[b]{0.48\textwidth}
         \centering
         \includegraphics[width=\textwidth]{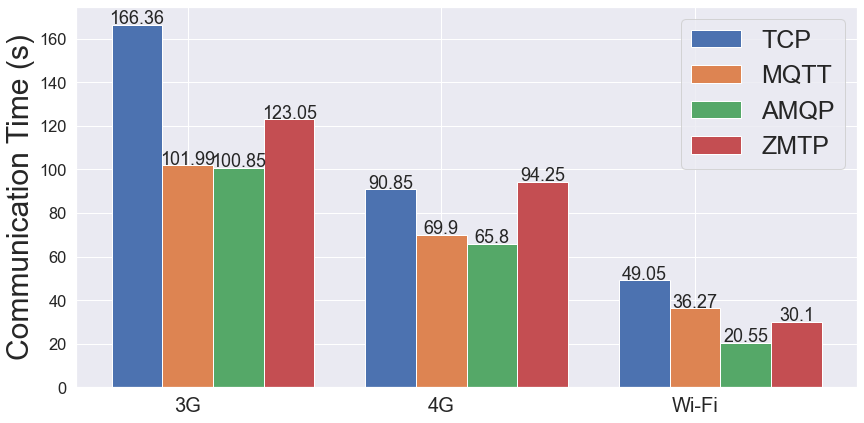}
        \caption{VGG-8}
    \label{fig:tcp_mqtt_amqp_net_vgg8}
     \end{subfigure}
         \caption{Communication time when using TCP socket, MQTT, AMQP and ZMTP after 5 training rounds of the VGG-5 and VGG-8 models for different network settings while under network stress.}
\end{figure}

\textbf{Communication time under network stress:} The effects of network stress are almost similar to those shown in Figure~\ref{fig:tcp_mqtt_amqp_comm_vgg5} and Figure~\ref{fig:tcp_mqtt_amqp_comm_vgg8}. This is because network stress works similarly to the \texttt{tc} command, limiting the network bandwidth. In Figure~\ref{fig:tcp_mqtt_amqp_net_vgg8}, TCP performs the worst with the slowest time and an increase in communication time of 158 seconds (Figure~\ref{fig:tcp_mqtt_amqp_comm_vgg8}). AMQP performs the best with the fastest time and an increase in communication time of 114 seconds (Figure~\ref{fig:tcp_mqtt_amqp_comm_vgg8}). MQTT performs moderately with an increase in communication time of 123 seconds (Figure~\ref{fig:tcp_mqtt_amqp_comm_vgg8}). MQTT and AMQP are designed to operate in poor network conditions because the server sends only one message to the broker, which is then distributed to all the devices. However, TCP sends the messages directly to the four devices that results in poorer performance by TCP. 

\begin{figure} [tp]
	\centering
	\includegraphics[width=0.48\textwidth]{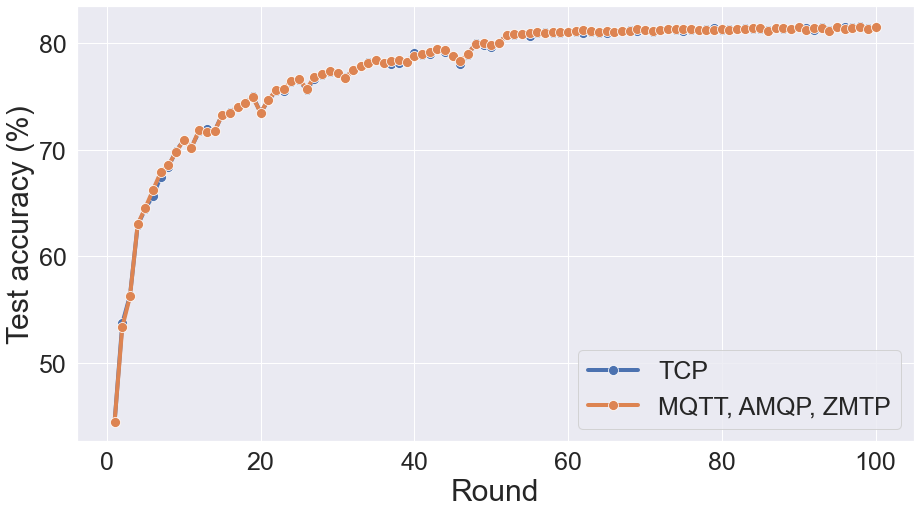}
	\caption{Accuracy when using TCP socket, MQTT, AMQP and ZMTP after 5 training rounds of the VGG-5 model under Wi-Fi conditions.}
\label{fig:tcp_mqtt_amqp_acc_vgg8}
\end{figure}

\textbf{Accuracy:}
The accuracy for socket TCP and all optimized application layer protocols remains largely similar with a few exceptions as in Figure~\ref{fig:tcp_mqtt_amqp_acc_vgg8}. The exceptions have a less than 0.5\% difference which may be due to how the \texttt{tc} command is used with the brokers or how the optimized application layer protocols have been designed. The main reason that accuracy is preserved is because these protocols all use TCP as the transport layer protocol.

\section{Discussion \& Conclusions}
\label{sec:discussion}

In this paper, we proposed \FedComm,\, a benchmarking method for FL application layer protocols. We evaluated the major application layer protocols that could be used in a FL context, such as standard TCP, standard UDP, MQTT, AMQP, and ZMTP. In summary, it is observed that the results of UDP seem inconsistent and not in line with the considerations taken into account when designing UDP. However, there are several factors related to the FL algorithm that may skew the UDP results. First of all, UDP requires additional computation, in sending the weights, requiring the model to be split up into multiple chunks that adhere to the UDP protocol size restriction. Furthermore, the UDP implementation uses TCP to send END messages to indicate the final weight sent per round; however, the implementation requires an additional thread to receive this, resulting in extra waiting periods and affecting communication time. Furthermore, if packets are lost during transmission, UDP requires additional computation to fill in the missing chunks and make sure the model is the correct size. This is common in poor network conditions; therefore, it would be computationally expensive to guarantee the missing chunks.

Optimized application layer protocols have outperformed the basic implementation of TCP/UDP in most aspects of the evaluation. According to the results, the two broker-based application layer protocols perform better, with AMQP having the best performance, followed by MQTT. Broker-based applications outperform socket-based implementations due to the messaging architecture and protocol design, despite the fact that broker-based applications may require more work to set up.

The ZMTP protocol uses the publish/subscribe model that broker-based protocols use, but it uses sockets to create one-way communication. This is less efficient than MQTT and AMQP, which affects the overall performance.

In the majority of experiments, the two socket-based implementations perform the worst. TCP socket implementation outperforms UDP in all experiments in terms of communication time and accuracy when network conditions are good, but performs the worst in terms of communication time when network conditions are poor. Therefore, when network conditions are not guaranteed, either AMQP or MQTT may be more appropriate for implementing FL as optimized application layer protocols. Moreover, the optimised application layer protocols demonstrated high accuracy with minor differences of less than 0.5\% at certain points. Therefore, when compared to non-optimized application layer protocols, optimized application layer protocols are more suitable for FL settings in both poor and good network conditions.




\textit{\textbf{Future Research:}} Currently, \FedComm\ evaluates a limited number of core application layer protocols and highlights the fundamental functionality of each protocol. As a future work, we highlight a number of application layer enhancements within the context of FL that will require further investigation.

\textit{Reliable implementation of socket UDP for FL:} We will improve and thoroughly test UDP socket implementation with additional parameters. \FedComm\ will support adjusting the size of UDP chunks, giving users more control over communication time and accuracy. 


\textit{Enhancement of standard MQTT, AMQP and ZMTP for FL:} MQTT will be evaluated using various broker configurations, including both local and public brokers, to provide a more in-depth analysis of real-world MQTT usage. Furthermore, MQTT for sensor networks will be added to \FedComm\ and compared to standard MQTT in the FL setting.

Different brokers will be implemented for AMQP to investigate the differences in how AMQP operates; will communication time be reduced? Will the accuracy improve? Will there be a change in packet loss? All of this will be thoroughly investigated. Moreover, we will attempt to change the architecture of AMQP and investigate how this different architecture will affect AMQP in a FL setting.

ZMTP is highly modifiable; various messaging types will be compared to all application layer protocols, including request/response, pull/push, pipeline, and pairs. To provide a more in-depth evaluation of the application layer protocols in various environments, we will attempt to scale the testbed by adding more devices.


\balance
\bibliographystyle{IEEEtran}  
\bibliography{references}

\begin{thebibliography}{10}
\bibitem{vineethfederated}
S.~Vineeth, ``Federated learning over wifi: Should we use tcp or udp?''
\end{thebibliography}

\end{document}